\documentstyle[11pt,newpasp,twoside,epsf]{article}
\def\edcomment#1{\iffalse\marginpar{\raggedright\sl#1\/}\else\relax\fi}
\reversemarginpar

\begin{document}
\title{Heating the Quiet Corona by Nanoflares: Evidence and Problems}
 \author{Arnold O. Benz}
\affil{Institute of Astronomy, ETH-Zentrum,
    CH-8092 Z\"urich, Switzerland }
\author{S\"am Krucker}
\affil{Space Sciences Laboratory, UCB, Berkeley CA 94720, USA}

\begin{abstract}
The content of coronal material in the quiet Sun is not 
constant as soft X-ray and high-temperature EUV line observations have 
shown. New material, probably heated and evaporated from the 
chromosphere is occasionally injected even in the faintest parts 
above the magnetic network cell interiors. Assuming that the 
smaller events follow the pattern of the well observed larger ones, we estimate the total 
energy input. Various recent analyses are compared and 
discussed. The results using similar EUV data from EIT/SOHO and TRACE basically agree on the power-law exponent when the same method is used. The most serious deviations are in the number of nanoflares per energy unit  and time unit. It may be explained at least partially by different thresholds for flare detection.
\end{abstract}

\section{Introduction}

In the past decade the resolution of quiet corona observations has constantly increased in space, even more in time, but most notably in flux. Long-exposure observations of the {\sl quiet} corona in soft X-rays by Yohkoh/SXT have revealed a large number of brighenings above the network of the magnetic field in quiet regions (Krucker et al. 1997). They have a typical thermal energy content of the order of $10^{26}$ erg and occur at a rate of 1200 events per hour over the whole Sun. More recently, the coronal emission measure in quiet regions has been observed in EUV iron lines with SOHO/EIT and was found to fluctuate locally at time scales of a few minutes in a large majority of pixels including the intracell regions (Benz \& Krucker 1998; Berghmans et al. 1998). At the level of 3 standard deviations, Krucker \& Benz (1998) reported the equivalent of 1.1$\times 10^6$ events per hour over the whole Sun and estimated the total energy input by the emission measure enhancements to be 16\% of the calculated total radiative loss of the observed region. The input estimate is limited by the sensitivity of the instrument, but also depends on some model parameters. In particular, the effective line of sight thickness of the coronal plasma (or height for observations in the center of the disk) cannot be measured and must be assumed. 

The distribution of the events in energy is also still controversial. Most observers report a power-law shape, but widely disagree in the exponent, which ranges from -1.45 (Berghmans \& Clette 1999, measuring radiative loss in Fe XII) and -2.59 (Krucker \& Benz 1998, measuring emission measure enhancements). Here we report agreement between their EIT based analysis and the studies by Parnell \& Jupp (2000) and Aschwanden et al. (2000) based on TRACE data, if the same method is used. For a flare model with a height given by the square root of the flare area, a simultaneous peak time within 2 minutes over the flare area and no further flare selection, all three investigations yield a power-law index in the range -2.0 and -2.4, the most likely new EIT value being -2.3. 

\section{Energy Estimates}

Brightness variations of individual pixels in the quiet corona are due to changes in emission measure, ${\cal M}:=\int n^2A\ ds$,  of the relevant line or, for the continuum, the relevant energy range of the detector. The integration is along the line of sight in $s$, $A$ is the pixel size, and the density $n$ refers to the plasma in the specified temperature range. As the temperature in the quiet corona is observed to be relatively stable between 1.0 and 1.6 $10^6$ K, the brightenings are not caused by changes in the temperature of the corona, but due to addition of new material. 

Energy estimates of heating events can be made from the radiation output or from the observed emission measure increase. In the former case, the line emission has to be integrated over the event and the total radiation estimated from some emissivity code. More problematic is the estimate of conduction losses that also take place during the whole time of enhanced emission. Estimates of energy losses are generally not accurate (Shimizu 1995). Thus we follow the latter way, estimating the input from the energy of the newly added material. Here some of the assumptions made along this estimate are discussed.

An increase in emission measure involves many forms of energies: heating of chromospheric material (thermal energy), lifting up the material to the corona (potential energy), expansion of the material from chromospheric density to coronal values, radiation and conduction losses during heating.

The thermal energy is $E_{\rm th} = 3 k_B T n_e V$ for the two particle species. The increase in emission measure $\Delta {\cal M}$ is directly observable, but requires some background subtraction. For $n_{e,0} \ll n_{e,1}$, high density material is injected into a small part of the old volume, and the thermal energy can be estimated as $E_{\rm th} \approx 3 k_B T\sqrt {\Delta {\cal M} A s_{\rm eff}}$ (cf. discussion in Brown et al. 2000).

The gravitational potential energy of a relatively large heating event has been estimated by Benz \& Krucker (1998). It amounts to 5.7\% of the thermal energy as estimated above.

For an expansion along a flux tube of constant diameter, the electron density decreases from the initial $n_{e,i}$ in the chromosphere to $n_{e,1}$ in the corona, where  $n_{e,i} \gg n_{e,1}$. The temperature is not seen to decrease from a very high value, thus the expansion may be modeled as being isothermal from an initial to a final length, $\ell_i$ and $\ell_1$. The expansion energy has been estimated by Benz \& Krucker (1999), who find a larger value than the thermal energy by a factor 2/3\ $\ln{(\ell_1/\ell_i)} > 1$.

Conduction and radiation losses occur during the rise phase of an event, and before the peak increase in emission measure is reached. Thus the thermal energy again underestimates the total energy input.

\begin{figure}
\plottwo{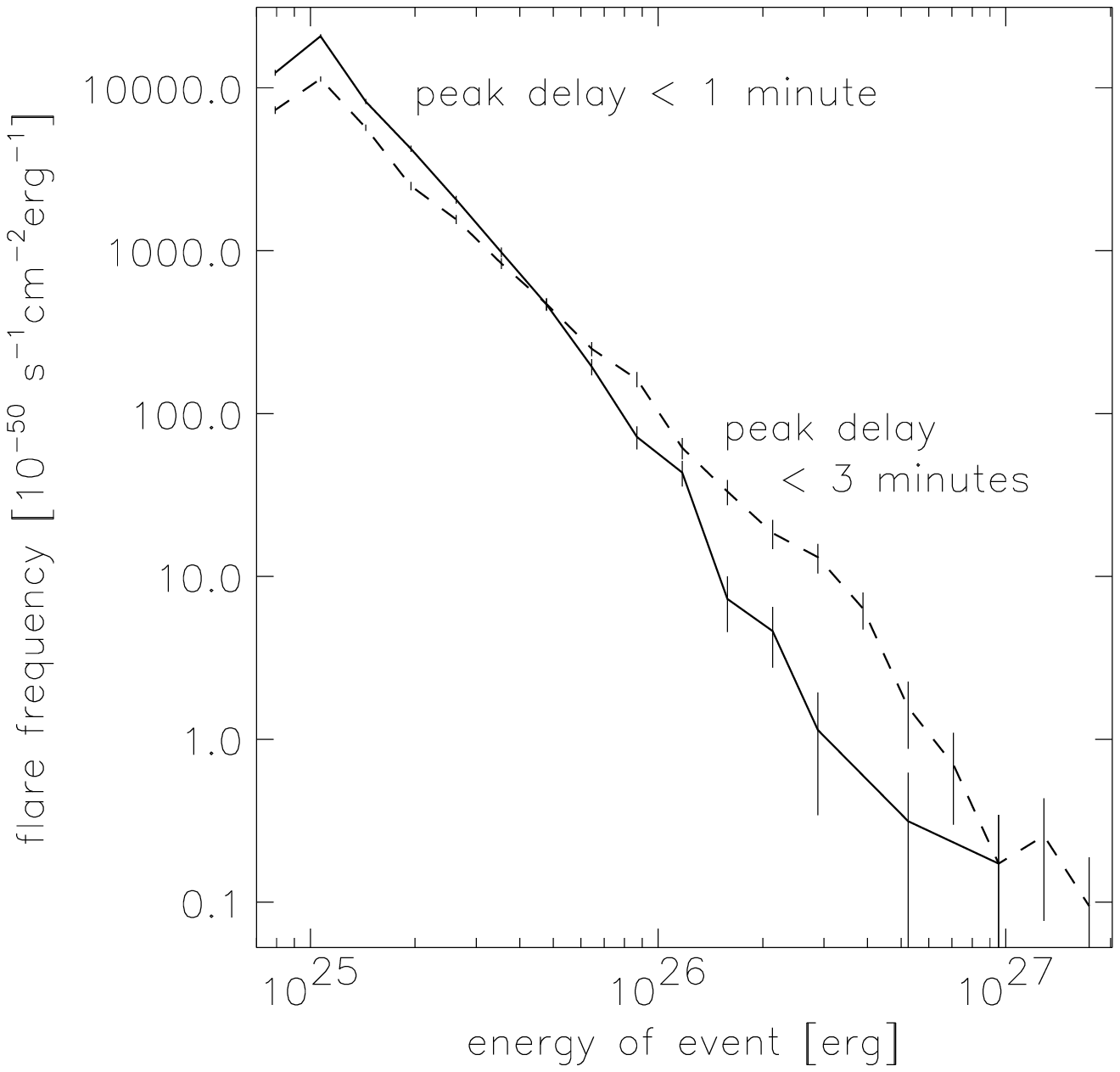}{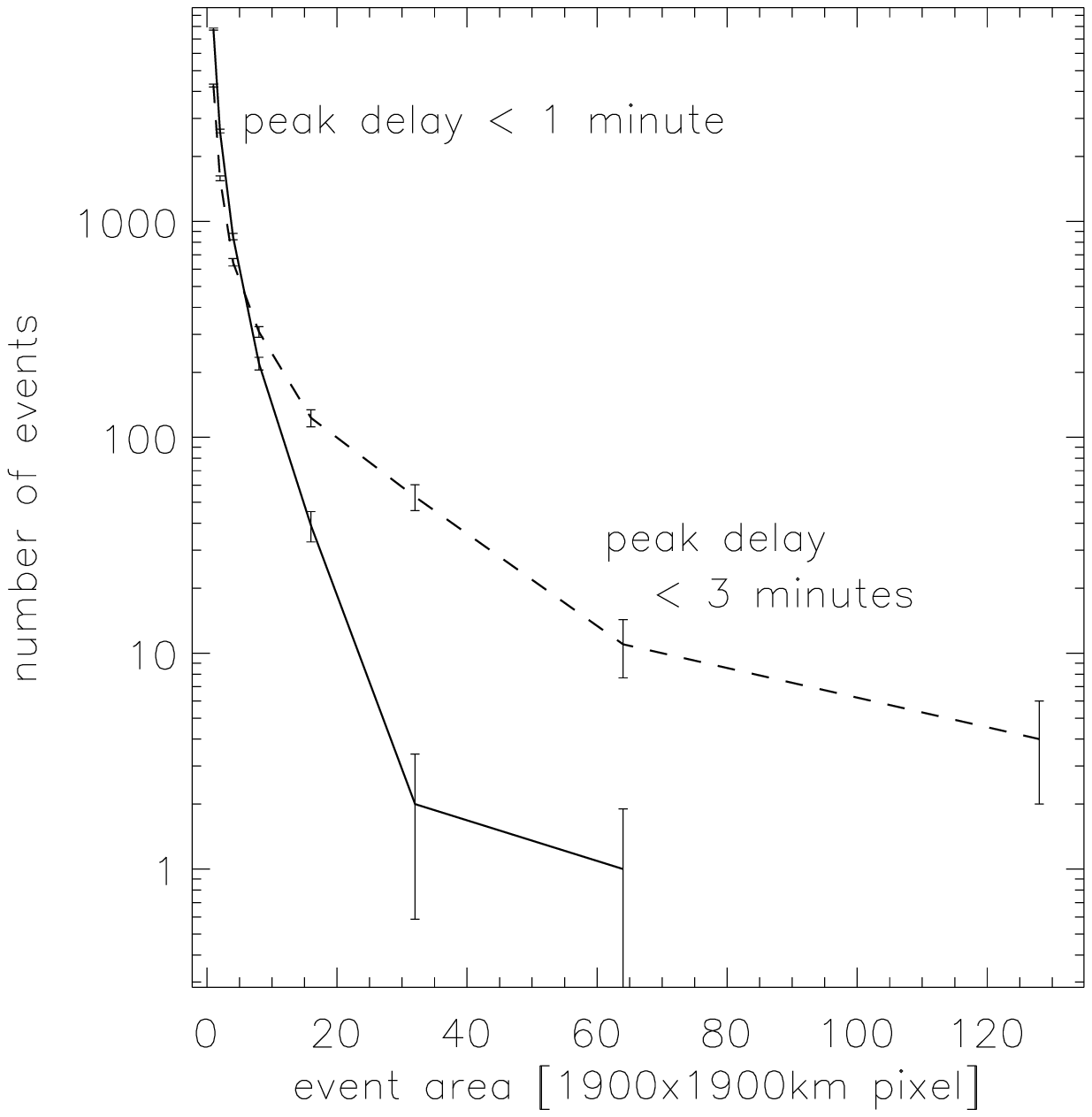}
\caption{{\sl Left:} The energy distribution of impulsive heating events for two different tolerances in the timing of adjacent pixel's peak.  The power-law index decreases from $2.59\pm0.02$ for the $\pm$1 minute requirement to $2.15\pm0.02$ for the $\pm$3 minute requirement for combining simultaneous events in adjacent pixels. {\sl Right:} The area distribution of impulsive heating events observed in the quiet corona. The number of very large events strongly increases from the $\pm$1 minute requirement to the $\pm$3 minute requirement for the simultaneity of adjacent peaks.}
\end{figure}

\section{Discussion and Conclusions}
To calculate the thermal energy of an event several assumptions have to be made. We have used the following simple procedure: The emission measure time series of each pixel have been searched for local peaks. If it exceeds the preceding minimum by more than 3 $\sigma$ (standard deviation of the noise as derived from the observed flux level), it is marked. Neighboring pixels peaking in the same time step (2 minutes) are combined to an event. The thermal energy of the event is calculated from the total emission measure increase. The method has been carefully tested and is described in Krucker \& Benz (1998).

In the following we discuss the robustness of the above result concerning the definition of events.  The most serious effect we found was in the tolerance of combining peaks in adjacent pixels into one event. Figure 1 (left) compares the condition for simultaneity of the peak times. If the condition is increased from $\pm$1 to $\pm$3 minutes, the number of large events increases at the expense of the small events. Therefore, the power-law exponent decreases. Increasing the tolerance may be justified if different parts of a source do not peak at the same time or by motions in the source as reported e.g. by Benz \& Krucker (1998). Parnell \& Jupp (2000) have found such motions to be rare in TRACE data. On the other hand, the larger tolerance also enhances the probability for chance associations, which is considerable at the observed rate of brightenings and in particular for events with large area. The combined events may then be artificially enhanced in area. This is shown in Fig. 1 (right) with the distribution of event size. The energy distribution becomes flatter as the event areas generally increase for larger tolerance. However, very large events, exceeding 4$\times 10^8$km$^{2}$, appear. We have found all of them to be unrealistic and most likely to be chance associations of more than one flare. Thus, the increase of the timing tolerance has a positive and a negative effect. The evaluation of these two effects and the correct combination of pixels to events need further investigations. 

Furthermore, sensitivity has important effects. The exponent changes from -2.59 to -2.39, when the $>$3$\sigma$ condition was enhanced to $>$6$\sigma$.  Most notably, the number of events at a given energy is reduced and the flare frequency is lower by a factor of 4.2. This latter effect originates from the higher sigma cutoff that eliminates some of the adjacent pixels with low-level variation. Thus the area of an event shrinks and so does the energy. The effect may explain the lower values for the flare frequency distribution observed in TRACE data (Parnell \& Jupp 2000; and some of the discrepancies of Aschwanden et al. 2000).

It is clear from EIT data that emission measure enhancements constitute a major energy input into the low, quiet corona. The enhancements have life times between 2 and more than 40 minutes (Berghmans et al. 1998), after which the plasma cools below the observational threshold. Thus the low corona, where most of the emission of the quiet Sun in soft X-rays and coronal lines originates and most of the heating must occur, is continuously supplied with newly heated material.
The emission measure enhancements resemble regular flares in active regions. The larger of these nanoflares in the quiet corona can be further investigated and have indeed revealed many similarities with regular flares (Krucker \& Benz 2000). Nanoflare heating of the quiet corona is well supported by the above investigations on total energy input and the relatively steep slope of the energy distribution.

\end{document}